\documentclass[a4paper,fleqn,usenatbib]{mnras}
\usepackage{newtxtext,newtxmath}
\usepackage[T1]{fontenc}
\usepackage{ae,aecompl}

\usepackage{graphicx}	
\usepackage{amsmath}	
\usepackage{amssymb}	
\usepackage{xspace}
\usepackage{multirow}
\usepackage{anyfontsize}

\newcommand{\kms}{kms$^{-1}$\xspace}	

\newcommand{\teff}{\ensuremath{T_{\text{eff}}}\xspace}
\newcommand{\rsun}{\text{R}\textsubscript{\sun}\xspace}
\newcommand{\msun}{\text{M}\textsubscript{\sun}\xspace}

\newcommand{\vsini}{$v\sin{i}$\xspace}
\newcommand{\radd}{$\text{rad~d}^{-1}$\xspace}
\newcommand{\s}{SS~Cyg\xspace}
\newcommand{\pc}{per~cent\xspace}
\newcommand{\ka}{\ensuremath{K_{1}}\xspace}
\newcommand{\kb}{\ensuremath{K_{2}}\xspace}
\newcommand{\ma}{\ensuremath{M_{1}}\xspace}
\newcommand{\mb}{\ensuremath{M_{2}}\xspace}
\newcommand{\ha}{\text{H}\ensuremath{\alpha}\xspace}
\newcommand{\hb}{\text{H}\ensuremath{\beta}\xspace}
\newcommand{\hg}{\text{H}\ensuremath{\gamma}\xspace}
\newcommand{\hd}{\text{H}\ensuremath{\delta}\xspace}

\title[The irradiated and spotted dwarf nova, SS~Cygni]{Roche tomography of cataclysmic variables - VIII: \\The irradiated and spotted dwarf nova, SS~Cygni}

\author[C. A. Hill et al.]{
C. A. Hill$^{1}$\thanks{E-mail: chill@irap.omp.eu}, Robert Connon Smith$^{2}$\thanks{E-mail: r.c.smith@sussex.ac.uk}, L. Hebb$^{3,4}$, P. Szkody$^{4}$\\
$^{1}$IRAP, Universit\'{e} de Toulouse, CNRS, UPS, CNES, 14 Avenue Edouard Belin, Toulouse, F-31400, France \\
$^{2}$Department of Physics \&\ Astronomy, University of Sussex, Falmer, Brighton BN1 9QH, UK \\
$^{3}$Hobart and William Smith Colleges, Department of Physics, Geneva, NY 14456, USA \\
$^{4}$Department of Astronomy, University of Washington, Box 351580, Seattle, WA 98195, USA}

\date{Accepted XXX. Received YYY; in original form ZZZ}

\pubyear{2017}

\begin{document}
\label{firstpage}
\pagerange{\pageref{firstpage}--\pageref{lastpage}}
\maketitle

\begin{abstract}
We present the results of our spectroscopic study of the dwarf nova \s, using Roche tomography to map the stellar surface and derive the system parameters. Given that this technique takes into account the inhomogeneous brightness distribution on the surface of the secondary star, our derived parameters are (in principle) the most robust yet found for this system. Furthermore, our surface maps reveal that the secondary star is highly spotted, with strongly asymmetric irradiation on the inner hemisphere. Moreover, by constructing Doppler tomograms of several Balmer emission lines, we find strong asymmetric emission from the irradiated secondary star, and an asymmetric accretion disc that exhibits spiral structures.

\end{abstract}


\section{Introduction}
\label{sec:intro}
First recognised by \cite{wells1896} as a variable star, and later classified as a dwarf nova (DN) by \cite{payne1938}, \s is one of the brightest cataclysmic variables (CVs) known, and is the brightest of the DN class. As is typical for CVs, \s consists of a white dwarf (WD) that accretes material from a main-sequence dwarf, overflowing its Roche lobe through the inner Lagrangian point. As the WD in \s is non-magnetic, accretion takes place from a boundary layer at the inner edge of the accretion disk, close to the WD surface. During quiescence, the low viscosity of the disk allows mass to accumulate within it. Once the surface density in the disk reaches a critical point, an instability is triggered, with an increase in viscosity allowing rapid mass flow through the disk to the WD, releasing gravitational potential energy \citep[the disk instability model; e.g.][]{warner1995}. The disk becomes heated, increasing in luminosity, and the system is observed in outburst, increasing in brightness from $V \simeq 12$ to around 8.5~mag during eruptions (lasting around 7--14~d), and returning to quiescence for around 50~d before the next outburst \citep{cannizzo1992}.

The system parameters of \s have been measured by many different authors over the past 50 years, and are now fairly well known, with the spectral type of the secondary star determined to be K4V--K5V \citep{smith1998, beuermann2000}. After the introduction of digital detectors, the radial-velocity semi-amplitude of the secondary star (\kb) was found to be around 155~\kms by \cite{stover1980}, as well as \cite{hessman1984}, \cite{robinson1986},  \cite{echevarria1989} and \cite{friend1990}. In later work, \cite{martinezpais1994} and \cite{bitner2007} found $\kb \simeq 162.5$~\kms, with \cite{north2002} finding a slightly higher value of 165~\kms. The differences between these measurements likely stemmed from real changes within the system, as an increase in irradiation of the secondary (due to increasing accretion luminosity) shifts the secondary's centre of light to higher velocities. This effect is most obvious when the system is in outburst (with an increase in \kb of $\sim40$~\kms, see \citealt{hessman1984}), but even during quiescent periods, there may be enough irradiation to significantly bias the measured \kb \citep{robinson1986}. Given that the mass ratio ${q = \mb/\ma = \ka/\kb}$ depends on accurate (unbiased) measurements, such variation due to irradiation will lead to systematically offset system parameters, if unaccounted for.

\begin{table}
\centering
\caption{APO echelle spectra; 600-s exposures}
\label{tab:spec}
\setlength\tabcolsep{6pt} 
\begin{tabular}{|l|c|c|}
\hline
Date & No. of spectra & Range of UTC mid-exposure \\
\hline
2012 Sep 20 & 29 & 04:13 to 10:36 \\
2012 Sep 21 & 29 & 02:29 to 09:07 \\
\hline
\end{tabular}
\end{table}

Furthermore, by fitting a two-spectrum model of a K4V and a cooler M2V component (that includes TiO in its spectrum), \cite{webb2002} found evidence of significant starspot coverage on the secondary star. Such surface features introduce significant radial velocity (RV) variations, and combined with the ellipsoidal variations due to the shape of the Roche lobe, RVs measured using standard cross-correlation techniques often depart significantly from their expected sinusoidal values (e.g. \citealt{hill2016}, also see Figure~\ref{fig:rvcurve}).

Moreover, as \s is non-eclipsing, determining the orbital inclination $i$ (and thus \ma and \mb) also presents some difficulties. One could assume some property of the K star, however this may introduce considerable uncertainties - \cite{kolb2001} have shown that Roche-lobe filling stars of K4V--K5V spectral type can have masses in the range 0.42--0.80~\msun, with \cite{baraffe2000} finding that mass-transferring stars may have radii up to 50~\pc larger than normal main-sequence stars, supported by observations of \cite{smith1998} and \cite{beuermann2000}. Instead, one could assume that the WD follows the normal mass-radius relation \citep[e.g.][]{provencal1998}, with the inner radius of the accretion disk allowing an inference of its mass \citep[e.g.][]{cowley1980}, and thus $i$. However, the velocity fields in \s are complex \citep{north2001}, especially near the WD where magnetic fields may play a role. More typically for non-eclipsing systems, one can determine $i$ by fitting the ellipsoidal variations in the lightcurve (e.g. \citealt{bitner2007}). However, this technique is also impaired by the presence of starspots and irradiation, complicated further by the fact one must estimate the flux dilution due to the accretion disk.

Thus, to accurately determine the system parameters for \s, one must account for inhomogeneities in surface brightness on the secondary star, as well as the ellipsoidal modulation, and variable contribution from the accretion disk. In this paper, we achieve this by using Roche tomography, a technique that reconstructs surface brightness maps, naturally accounting for the aforementioned effects, and thus (in principle) providing the most robust estimate of the system parameters, as well as allowing us to study starspot distributions and irradiation patterns.

We present our photometric and spectroscopic observations in Section~\ref{sec:obs}, and derive a new ephemeris from our RV measurements in Section~\ref{sec:ephemeris}. We examine the emission lines using Doppler tomography in Section~\ref{sec:emission}, and determine the system parameters using Roche tomography in Section~\ref{sec:roche}. In Section~\ref{sec:surfacemaps}, we analyse the surface maps of \s that show both irradiation and starspots, and discuss and summarize our results in Section~\ref{sec:discussion}.



\section{Observations}
\label{sec:obs}
\subsection{Spectroscopy}
\label{sec:spectroscopy}

\begin{figure}
\includegraphics[width=\columnwidth]{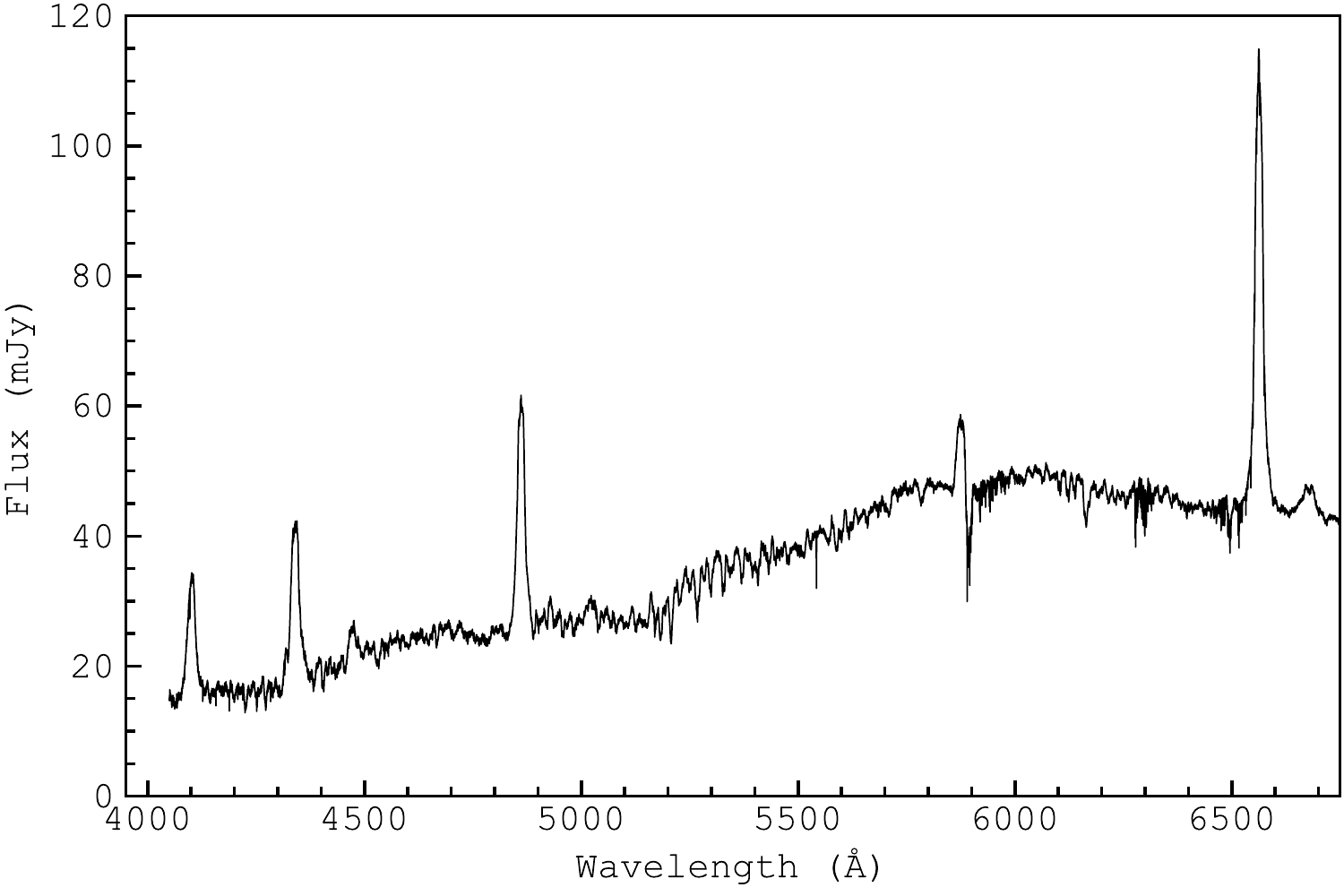}
\caption{A section of the average spectrum of SS Cygni from the first night's observing, showing the strong Balmer emission lines and some of the many absorption lines that were available for mapping (Section~\ref{sec:roche}).}
\label{fig:spectroscopy}
\end{figure}

The spectra were obtained in September 2012 using the high-resolution (R$\sim31,500$) echelle spectrograph on the ARC 3.5-m telescope at the Apache Point Observatory (APO), New Mexico. Each exposure was for 600\,s. A summary is given in Table~\ref{tab:spec}; approximately one orbital cycle was covered on each night. The echelle spectra were reduced to 1-D using standard reduction techniques, and all orders were placed on the same uniform wavelength scale of 0.1\,\AA\ per pixel and stitched together to form single spectra. These spectra were then clipped at each end, resulting in final spectra running from 4000--9000~\AA\ inclusive. A section of the average spectrum of the first night's data is shown in Figure~\ref{fig:spectroscopy}. Regions of major atmospheric molecular absorption (oxygen and water bands) were set to a constant and masked out when using the spectra. A flux-standard star (EG 247 = BD+52$^{\rm{o}}$\,913) was observed and reduced in the same way, and was used for flux calibration of the echelle spectra. 


%
%

\subsection{Photometry}
\label{sec:photometry}

\begin{figure}
\includegraphics[width=\columnwidth]{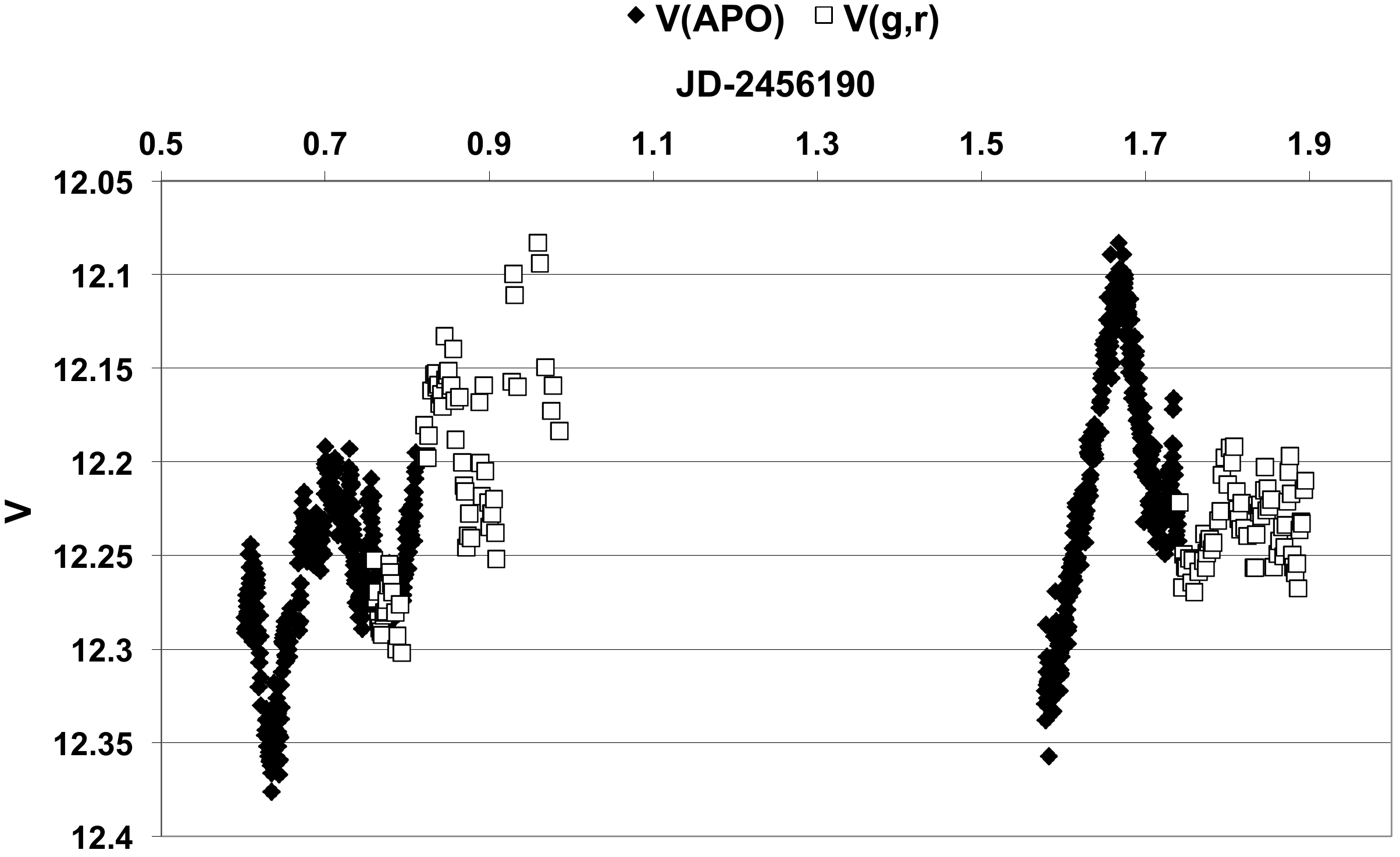}
\caption{Photometry as a function of time over the two nights. Filled diamonds mark the $V$-magnitude data from the APO; open squares mark the pseudo-$V$-magnitude data using equation \ref{transformation}, shifted to match the APO data in the region of overlap.}
\label{fig:UTplot}
\end{figure}

\begin{figure}
\includegraphics[width=\columnwidth]{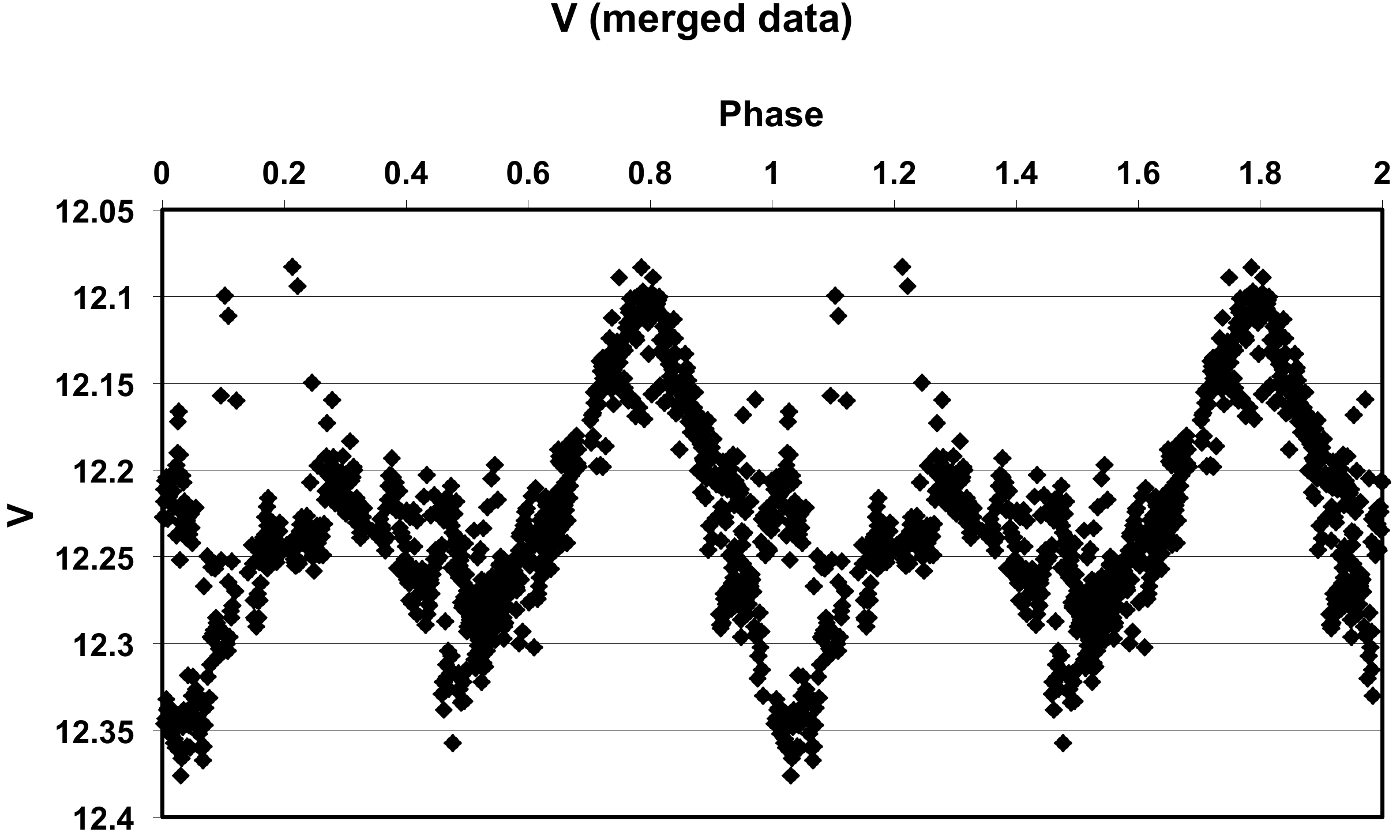}
\caption{Plot of the merged data from both nights as a function of phase, over two phase cycles. There are some obvious outliers, almost all from the pseudo-data. The scatter around phase 0 and phase 0.4 arise from variations between the two nights (see Figure~\ref{fig:UTplot}).}
\label{fig:Phaseplot}
\end{figure}

Simultaneous photometry was obtained on both nights, using two instruments. The overall brightness level was consistent with the system being in quiescence, and AAVSO data for the time confirm that our observations were obtained in quiescence, almost exactly half-way between two outbursts. Single-colour $V$ photometry was obtained on the NMSU 1-m telescope at APO; because that did not cover the entire timespan of the spectroscopy, it was supplemented with multi-colour ($g, r, i$) photometry taken at the University of Washington 0.76-m at Manastash Ridge Observatory, Washington. Because the two sets of photometry overlapped in time on each night, it was possible to combine them, using a transformation to create a pseudo-$V$ magnitude from a combination of the $g$ and $r$ data, using the formula:
\begin{equation}
V = g - 0.03 - 0.42(g-r)
\label{transformation}
\end{equation}

\noindent \citep[see][]{windhorst1991,Kent1985}. Figure \ref{fig:UTplot} shows the data for the two nights, distinguishing between the APO data and the ($g$,$r$) data; the latter show considerable scatter on night 1. The two nights show somewhat different behaviour, which shows up as scatter in the phase plot (Figure~\ref{fig:Phaseplot}), which combines the two nights and the two sources of data. The obvious outliers mainly correspond to the considerable scatter in the light curve at the end of night 1 in Figure~\ref{fig:UTplot}.

The mean $V$ band magnitude of $m_{v} = 12.23$ (as determined from our observations) is somewhat fainter than the mean historical quiescent magnitude of $m_{v} = $11.8--11.9 (between 1896 and 1992, \citealt{cannizzo1992}), which may imply a lower accretion rate (and corresponding accretion luminosity) at the time of our observations. Combining our mean $m_{v}$ with the fractional flux contribution of the secondary of $\sim 46.5$~\pc in the $V$ band \citep[as found by][]{bitner2007}, we derive a magnitude of $m_{v} = 13.064$ for the secondary star. Adopting a distance of $d = 114\pm2$~pc, as measured using VLBI by \citet{millerjones2013}, and in excellent agreement with the $116.8\pm4.5$~pc found by Gaia \citep{gaia2016}, we derive a distance modulus of $5.28\pm0.04$. Combining this with the bolometric correction of $BC_{v} = -0.55$ (for a K4V-type star), as determined by \citet{pecaut2013}, and assuming negligible extinction ($A_{v} \sim 0$), we derive an absolute bolometric magnitude of $6.2\pm0.1$ for the secondary star, or equivalently, a log luminosity relative to the Sun of $-0.58\pm0.04$.



\section{Ephemeris and Radial velocity curves}
\label{sec:ephemeris}
\begin{figure}
\includegraphics[width=\columnwidth]{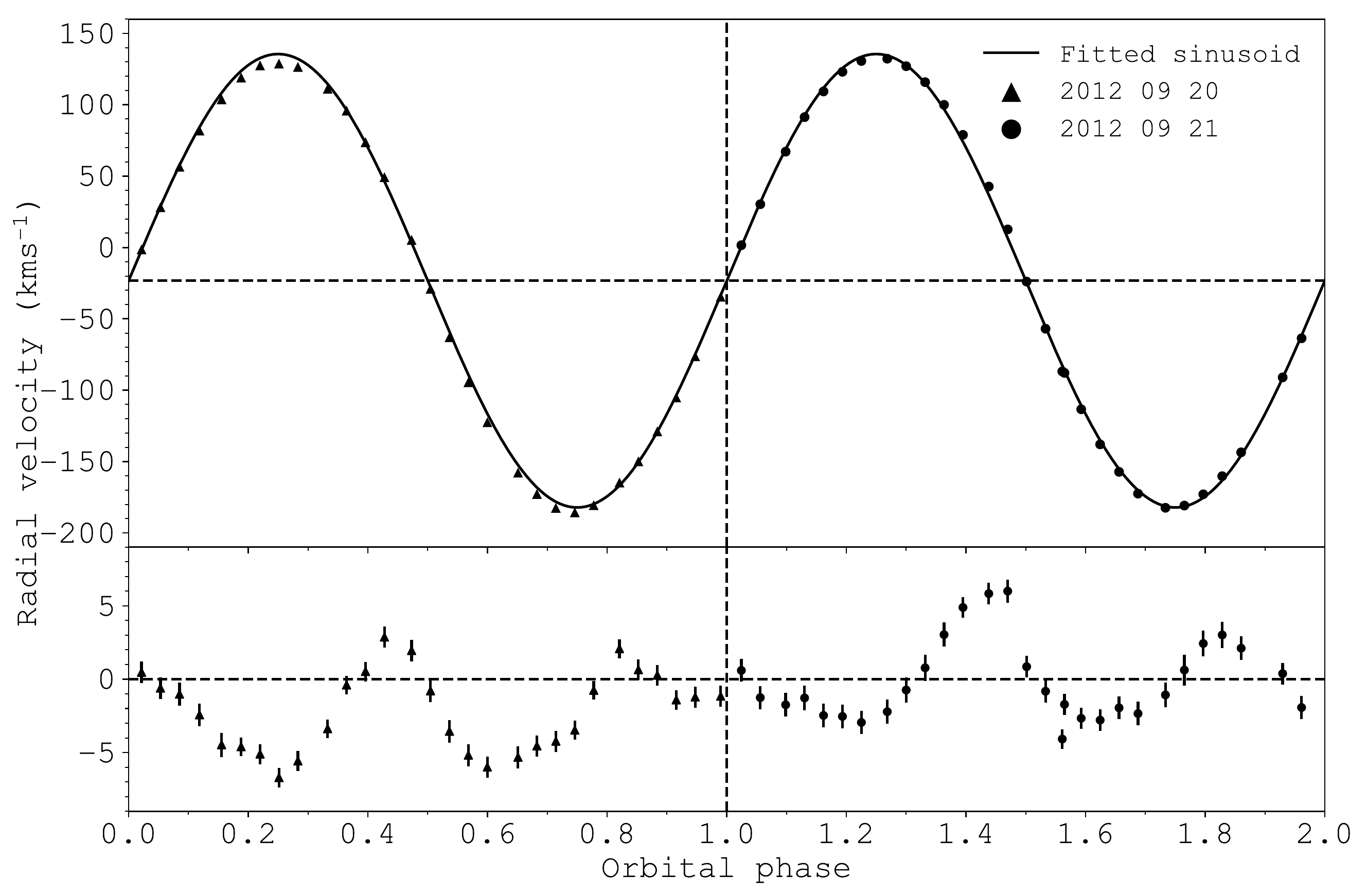}
\caption{The measured radial velocities of \s (top panel) for 20 Sept 2012 (triangles) and 21 Sept 2012 ( circles), where the latter points have been phase folded for clarity. Also shown is a least-squares sinusoid fit to the RV points, assuming a circular orbit (black solid line). The lower panel shows the residuals after subtracting the fitted sinusoid, as well as the statistical uncertainties of the measured RVs.}
\label{fig:rvcurve}
\end{figure}

A new ephemeris was derived for \s for the purpose of improving the quality of the Roche tomograms in Section~\ref{sec:roche}. Following the methods outlined by \cite{watson2006}, the radial velocities (RVs) were obtained by cross-correlating the observed spectra with a rotationally-broadened spectral-type template, over the range 6000--6270~\r{A} and 6320--6500~\r{A} (a region containing strong absorption lines, excluding tellurics). Our chosen K4V template was HD~24916, where its RV of $-5.144\pm0.005$~\kms was determined by a Gaussian fit to a least-squares deconvolved line profile (LSD, see Section~\ref{sec:roche}). The measured RVs were determined from parabolic fits to the peaks of the cross-correlation function for each spectrum, with a new zero-point for the ephemeris (Equation~\ref{eq:ephemeris}):
\begin{align}
\text{HJD (d)} &= 2456190.62770\pm0.00004 + 0.27512973 \text{E},
\label{eq:ephemeris}
\end{align} derived from a least-squares sinusoidal fit to the measured RVs (top panel of Figure~\ref{fig:rvcurve}). For all work presented here, we adopted the fixed orbital period of 0.27512973~d as found by \cite{hessman1984}.

This method of measuring the RVs is relatively insensitive to the use of a poorly-matched template, or an incorrect amount of template broadening. Indeed, no detailed attempt was made to determine the best-fitting spectral-type or binary parameters in this analysis. However, for completeness we find a systemic velocity of ${\gamma = -23.3\pm0.1}$~\kms and a radial velocity semi-amplitude ${\kb = 158.86\pm0.14}$~\kms (shown with their statistical uncertainties), with a projected rotational velocity \vsini = 93.6~\kms. While \kb and \vsini are compatible with those derived from our tomographic imaging (see Section~\ref{sec:roche}), the value of $\gamma$ is around 8~\kms lower. This difference likely stems from the non-uniform intensity distribution on the inner hemisphere of the secondary star, as seen in Figure~\ref{fig:map}. Such a phenomenon was also found for HU~Aqr \citep{watson2003}, where a non-uniform intensity distribution, caused by the accretion curtain and disk partially blocking irradiation, was found to account for a 14~\kms velocity shift in the measured value of $\gamma$. We note that the values of $\gamma$ and \kb also change significantly when \s is in outburst \citep[see][]{hessman1984}; however, the system was in quiescence during the observations presented here (see Section~\ref{sec:photometry}), and so we consider the surface brightness inhomogeneities to be the source of this discrepancy.

The lower panel of Figure~\ref{fig:rvcurve} shows the residuals after subtracting the fitted sinusoid. Here we see the systematic biases in the measured RVs due to surface features such as irradiation or star spots, as well as the tidal distortion of the secondary. Given that we are able to account for such phenomena by using Roche tomography (see Section~\ref{sec:roche}), the binary parameters derived from this RV analysis have not been used in the subsequent work presented here.

\section{Emission line features}
\label{sec:emission}
\begin{figure}
\includegraphics[width=\columnwidth]{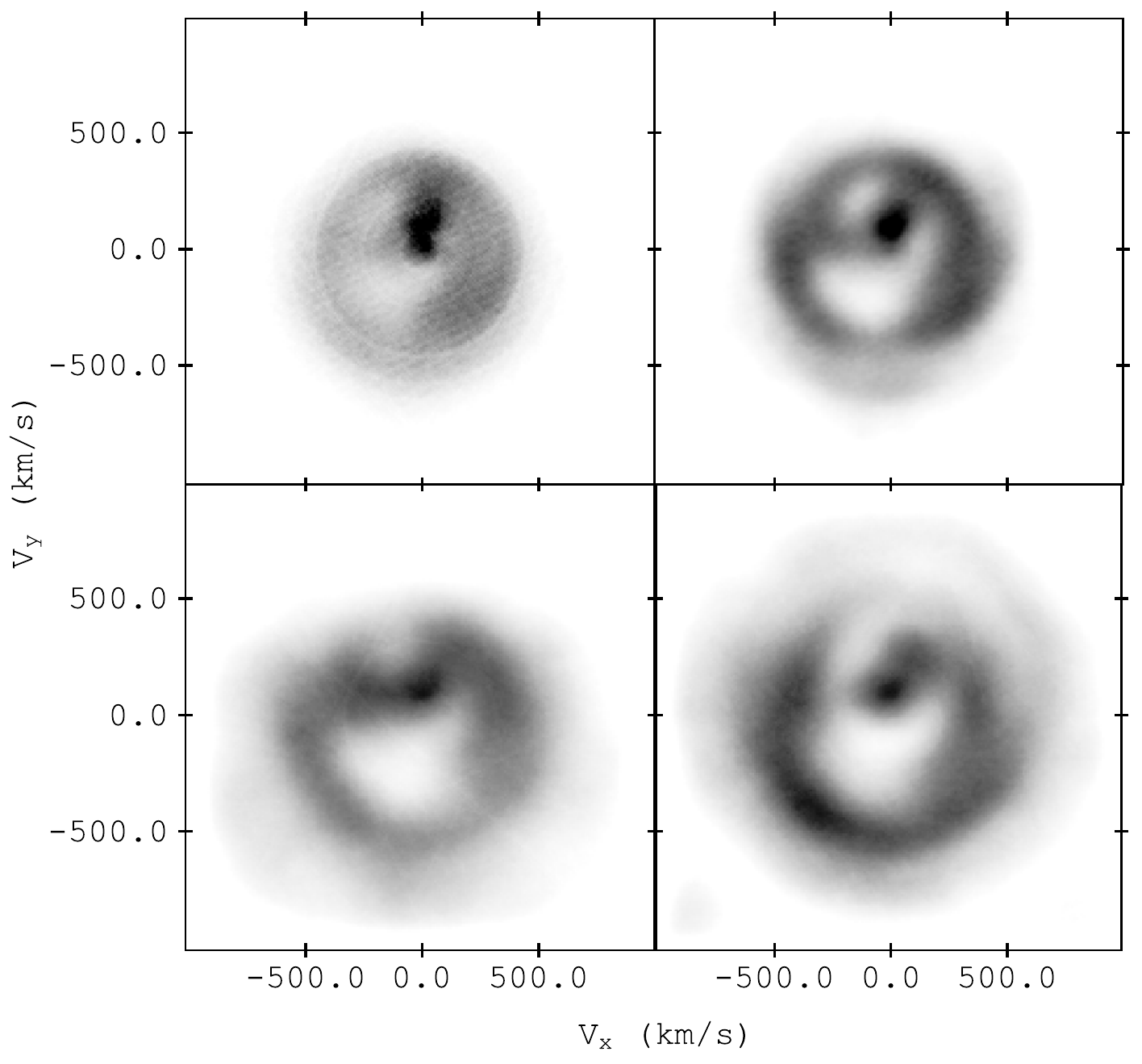}
\caption{Doppler tomograms of the \ha, \hb, \hg and \hd emission lines in \s (in clockwise order from top left). A spiral structure can be clearly seen in the tomograms of \hb, \hg and \hd, as well as the accretion stream in \ha, \hb and \hg.}
\label{fig:doppler}
\end{figure}

The Balmer emission lines in \s are strong, and it is of interest to discover whether they arise entirely in the accretion disc or partly in the irradiated hemisphere of the secondary star. To explore this we have constructed Doppler maps from the \ha, \hb, \hg and \hd emission lines, shown in Figure~\ref{fig:doppler}.

In the case of \ha and \hb, we see that a large portion of the emission comes from the secondary star, with a slight asymmetry in the intensity distribution towards positive velocities. Furthermore, the tomograms of \ha, \hb and \hd show the accretion stream along its expected ballistic trajectory. We see spiral structures in the disc in the tomograms of all emission lines, with the spiral arms overlapping somewhat in the lower-right quadrant of the maps. Indeed, our observations are consistent with those by \citet{kononov2012}, who found a similar asymmetric emission in the right-hand quadrants. We also find increased emission in the lower-left quadrant of the \hg map, similar to that found in the \ha tomogram of \citet{north2002}, as well as the \hg and \hd tomograms of \citet{kononov2012}. However, there are also differences between our tomograms and those of \citet{kononov2012}, confirming their conclusions of changes in the accretion disk between outbursts.

\section{Roche tomography}
\label{sec:roche}
We have used Roche tomography to map the surface of the secondary star in \s, and derive the system parameters. Analogous to Doppler imaging, this technique is specifically designed to indirectly image the secondary stars in close binaries such as CVs, and has been successfully applied to many systems over the past 20 years \citep[e.g.][]{rutten1994,schwope2004,watson2003,watson2006,watson2007b,dunford2012,hill2014,hill2016,parsons2016}. The technique assumes that the secondary is locked in synchronous rotation with a circularized orbit, and that the star is Roche-lobe filling. Surface brightness maps are reconstructed from a time-series of spectra using a maximum-entropy approach, where the least amount of information is included to reproduce the observed line profiles, for a given $\chi^{2}$. We refer the reader to the references above and the technical reviews of Roche tomography by \cite{watson2001} and \cite{dhillon2001} for a detailed description of the axioms and methodology.

\begin{figure*}
\includegraphics[width=0.8\textwidth]{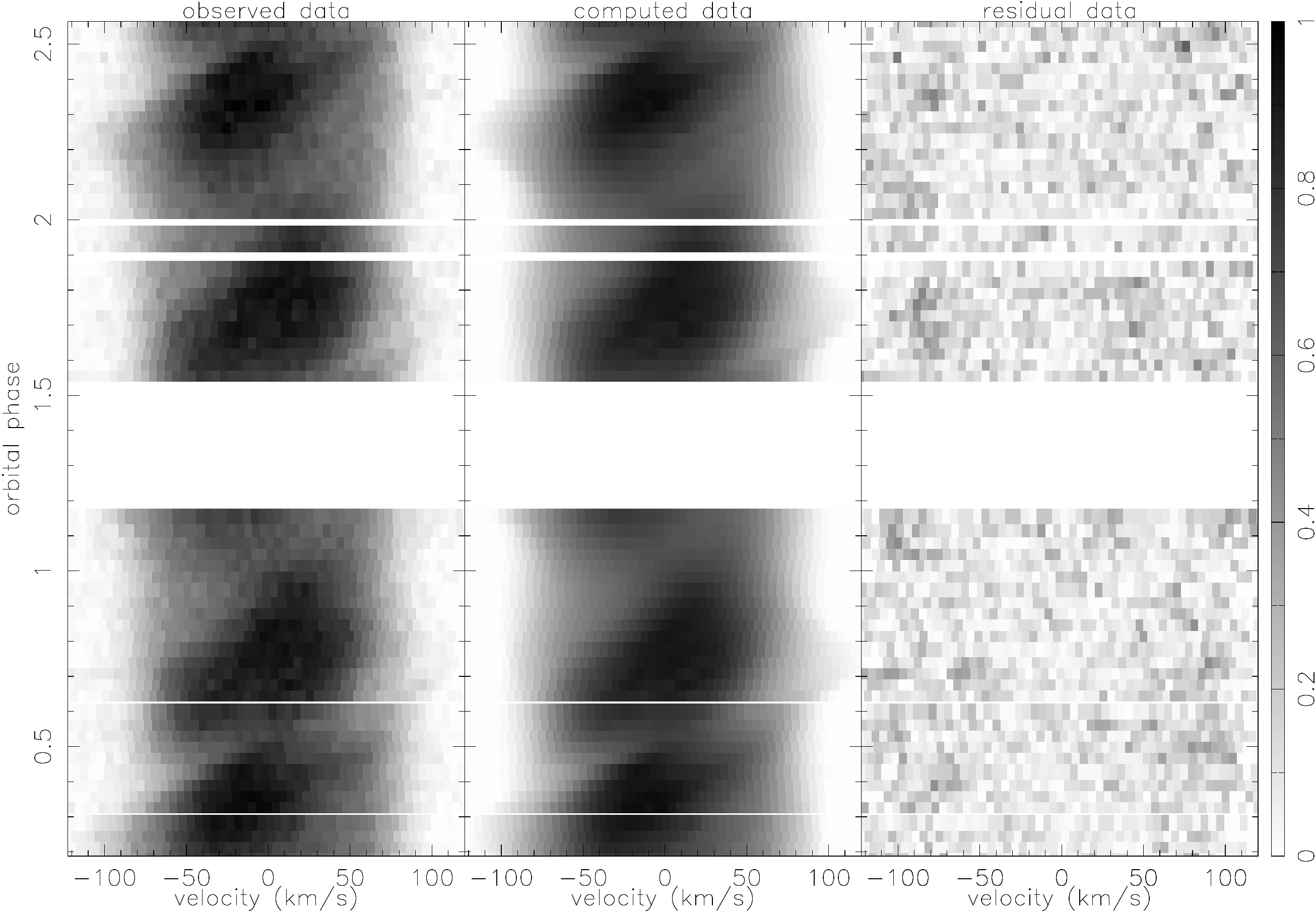}
\caption{Trailed LSD profiles of SS~Cyg, where the orbital motion has been removed assuming the binary parameters found in Section~\ref{sec:systempars}, allowing individual starspot tracks across the profiles and the variation in \vsini to be more clearly observed. Panels show (from left to right) the observed LSD profiles, the computed fits to the data using Roche tomography, and the residuals (increased by a factor of 5). Starspots and surface features appear bright in these panels, where a grey-scale of 1 corresponds to the maximum line depth in the reconstructed profiles. The two nights are displayed separately.}
\label{fig:trail}
\end{figure*}

In order to detect the small variations in line profile shape due to surface features (typically less than 10~\pc of the line depth), we applied least squares deconvolution (LSD; see \citealt{donati1997}) to all spectra, in the same manner as \cite{hill2014}. This cross-correlation process assumes all lines included in the LSD mask repeat the same information, and effectively yields a `mean' line profile with a substantially increased signal-to-noise ratio. LSD requires the spectral continuum to be flattened, but as the contribution from the accretion regions to each spectrum is unknown, with a constantly changing continuum slope due to, for example, flaring (see Figure~\ref{fig:UTplot}) or the varying aspect of the accretion regions, we cannot use a master continuum fit to the data. Furthermore, the constantly varying contribution of the secondary star to the total system light means normalizing the continuum would cause photospheric absorption lines from the secondary to vary between exposures. Thus, the continuum is subtracted from each spectrum using a spline fit. Given the spectral type of \s has been determined to lie in the range K4V--K5V \citep{smith1998}, we generated a stellar line list for a K4V type star ($\teff = 4750$~K and $\log{g} = 4.5$, the closest approximation available) using the Vienna Atomic Line Database \citep[VALD;][]{ryabchikova2015}, adopting a detection limit of 0.1 (of the normalized line depth, below which all lines with a smaller central depth were excluded). The normalized line depths were scaled by a fit to the continuum of a K4V template star so each line's relative depth was correct for use with the continuum subtracted spectra. Regions of the spectrum containing strong atmospheric telluric lines, and emission lines from accretion regions (such as Balmer lines) were excluded, leaving around 7500 lines for use in the LSD mask.

The resulting LSD profiles can be seen in Figure~\ref{fig:trail}, where starspots and irradiation both appear as apparent emission bumps.

The effects of limb darkening were accounted for during the fitting process by adopting the four-parameter non-linear limb darkening model of \cite{claret2011}, where the coefficients were taken from \cite{claret2012}. By using the stellar parameters closest to that of a K4V star, which for the \textsc{phoenix} model atmosphere were $\log{g} = 4.5$ and $\teff = 4800$~K, we derive coefficients of $a_{1} = 3.05$, $a_{2} = -6.42$, $a_{3} = 7.21$, $a_{4} = -2.51$. In addition, we accounted for gravity darkening by utilizing the expression presented by \cite{bloemen2011} and the coefficients given by \cite{claret2012}, where we interpolated between the Johnson B and V passbands using the effective central wavelength of the LSD mask, yielding a gravity darkening coefficient of $y(\lambda) = 0.76$.

\subsection{System parameters}
\label{sec:systempars}
\begin{figure}
\includegraphics[width=\columnwidth]{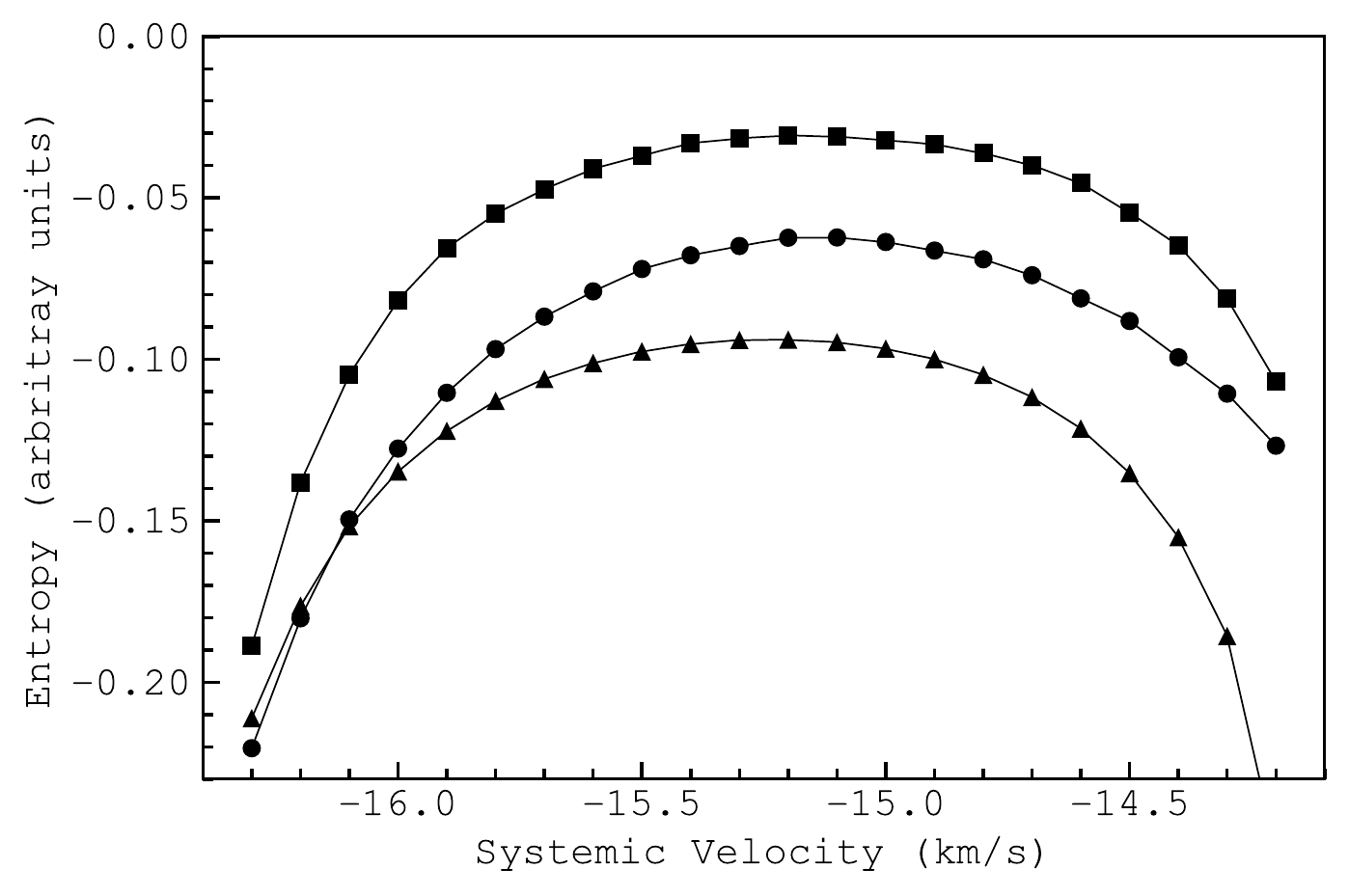}
\caption{The map entropy as a function of systemic velocity for the first night's data (triangles), the second night's data (circles) and the combined data (squares). Spline fits are shown only as a visual aid.}
\label{fig:systemic}
\end{figure}

\begin{figure}
\includegraphics[width=\columnwidth]{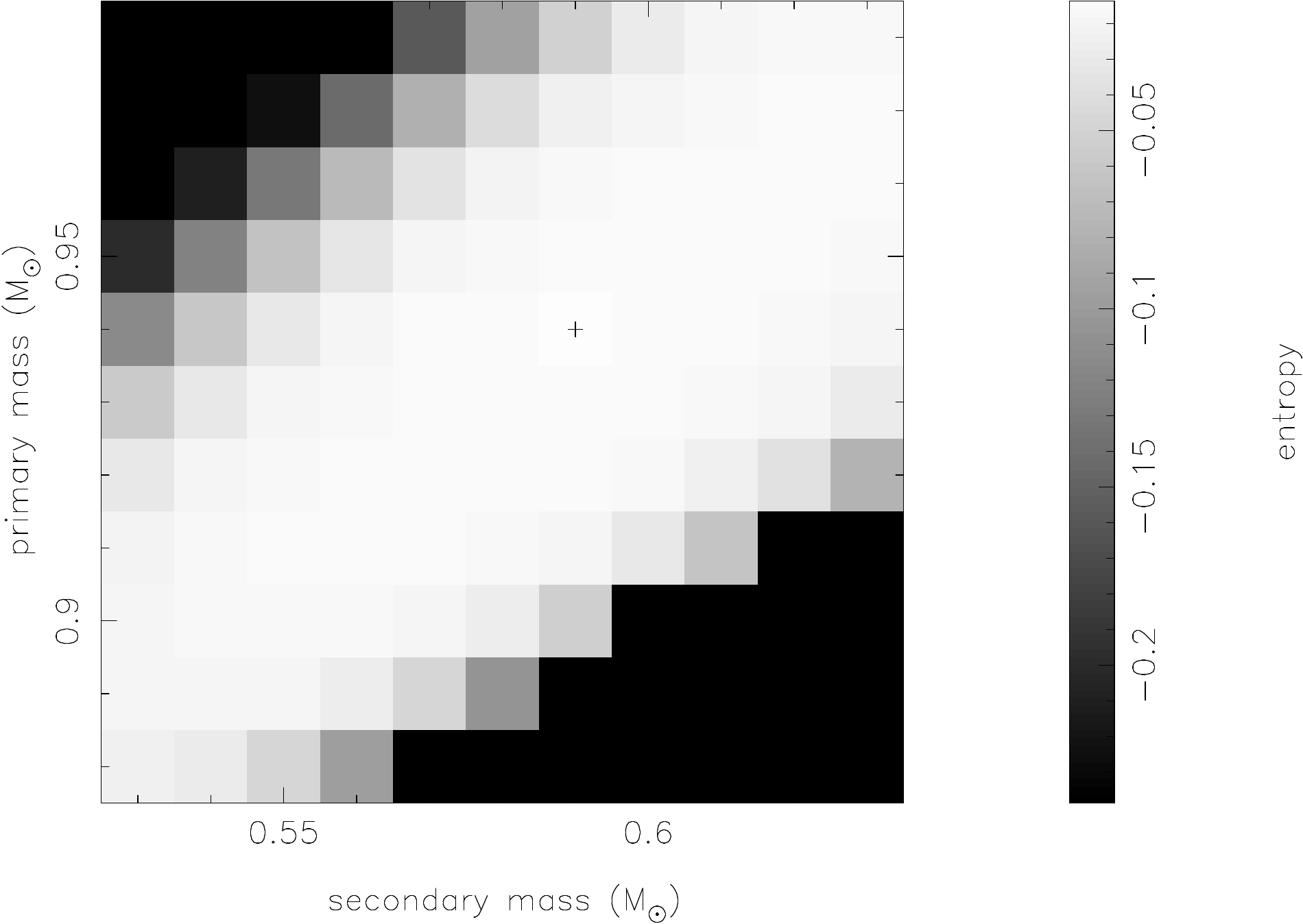}
\caption{An example of an `entropy landscape' at an inclination of $i=45\degr$, where Roche tomograms are reconstructed for many pairs of primary and secondary star masses, with the optimal pair of masses taken as those that produce the map of maximum entropy.}
\label{fig:entland}
\end{figure}

When using Roche tomography, the system parameters ($\gamma$, $i$, \ma and \mb) are typically determined by fitting the observed line profiles to the same level of $\chi^{2}$ for many combinations of parameters, aiming to minimize the information content of the reconstructed maps. Adopting incorrect parameters leads to well-characterized artefacts in the final map (see \citealt{watson2001}), and always acts to increase the information content (i.e. decreasing the map entropy). The fitting of data is carried out iteratively, varying each parameter separately until the map of maximum-entropy is obtained. In practice, the systemic velocity $\gamma$ is most easily constrained (see Figure~\ref{fig:systemic}), and the optimal pair of \ma and \mb are determined for each inclination $i$ (in an `entropy landscape', see Figure~\ref{fig:entland}). Thus, the optimal system parameters are those that produce the map containing least informational content (i.e. those  that assume least about the shape of the line profile), for a target $\chi^2$. 

For \s, we fitted the two nights' data separately, allowing for independent measurements of the system parameters, as well as creating independent surface maps (see Appendix~\ref{app:a}). It was found that the same set of parameters provided the optimal fit to each night's data. Furthermore, when the two nights' data were combined, we again obtained the same set of optimal parameters, showing the robustness of the technique against systematic biases (such as using different phase coverage or fewer spectra), and providing confidence in our derived results. Our fits to the LSD profiles, with residuals, are shown in Figure~\ref{fig:trail}. We find the systemic velocity ${\gamma = -15.2}$~\kms, the inclination ${i = 45\degr}$, the primary mass ${\ma = 0.94}$~\msun and the secondary mass ${\mb = 0.59}$~\msun (with a mass ratio of $q = \mb/\ma = 0.628$). Furthermore, we determine \vsini to range between a minimum of 88.7~\kms (at phase 0.0 and 0.5) and a maximum of 108.8~\kms (at phase 0.25 and 0.75), with \kb equal to 163.9~\kms. 

Using our derived masses, we find a volume-averaged radius of 0.696~\rsun \citep{eggleton1983}. Comparing this to the evolutionary models of \cite{siess2000} (where the radius of a 0.6~\msun main sequence star is around 0.53~\rsun), and the evolutionary models of \citet{baraffe2000} (where a 0.6~\msun main sequence star has a radius of around 0.57~\rsun), we find that the radius of the secondary star in \s is around 20--30~\pc larger than that of an equivalent main sequence dwarf.

\section{Surface maps}
\label{sec:surfacemaps}

\begin{figure}
\includegraphics[width=\columnwidth]{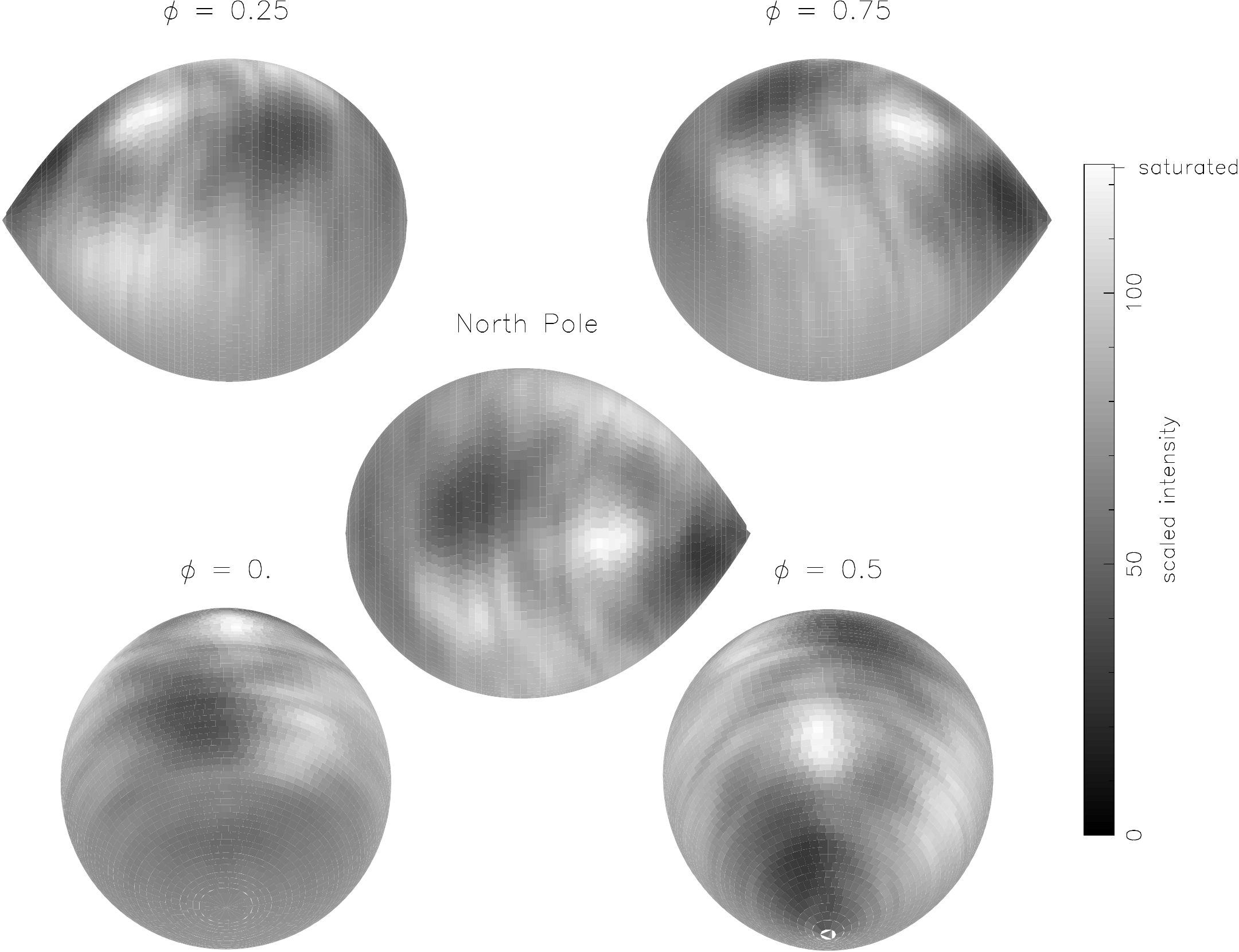}
\caption{The Roche tomogram of \s using both nights' data. Dark grey-scales indicate regions of reduced absorption line strength that is due to either the presence of starspots or the impact of irradiation. The orbital phase is indicated above each panel, and the tomogram is shown without limb darkening for clarity.}
\label{fig:map}
\end{figure}

Using the parameters derived in Section~\ref{sec:systempars}, we constructed Roche tomograms for each night's data (shown in Appendix~\ref{app:a}). Given that the data sets were taken on sequential nights, we are able to compare directly the reconstructed surface features for short-term evolution, as well as consistency in the reconstruction technique itself. Starspots (showing as dark features in the maps) are not expected to evolve significantly over such short timespans, typically taking weeks to months to emerge or disappear \citep[e.g.][]{parsons2016, isik2007, schrijver2002, cameron1995}. Furthermore, any longitudinal shear due to differential rotation should be minimal, given the measured shear rate of $d\Omega = 0.0233$~\radd for the CV, AE~Aqr \citep{hill2014}. Indeed, we may expect the most significant variability to occur in the level of irradiation from the white dwarf and accretion regions, leading to variations in the irradiation pattern on the inner hemisphere (showing as dark regions in the maps). However, we are generally insensitive to short-term variations in irradiation (due to, e.g., variations in accretion luminosity), as the effects are time-averaged over the duration of the data collection. We find that the reconstructed surface features are very consistent between maps, showing no significant morphology change, other than what may be expected by using data of different S/N, with different phase coverage. In particular, the long narrow spot feature extending from the pole to the equator around phase 0.75 is reconstructed in a similar manner in both maps, showing that even small structures have been reliably mapped, with $1\sigma$ brightness variations of less than 5~\pc between maps.

Given that we see no significant difference between the two nights' maps (see Figure~\ref{fig:maps}), we combined all the data to create a single map (shown in Figure~\ref{fig:map}), on which all subsequent analysis is based.

We find a large high-latitude spot (centred around $70\degr$ latitude) on the rear of the star, as is typically seen in many rapidly rotating single and binary stars \citep[e.g.][]{donati1999,barnes2000,hussain2007,watson2007b,shahbaz2014}. In addition, we find many smaller spots across the stellar surface, with some long and narrow features extending from high to low latitudes (in particular around $\phi = 0.15$ and 0.8). Such narrow features may be artefacts due to flaring, phase-undersampling, or may be a result of the `mirroring' effect \citep[that acts to smear features in the latitudinal direction, see][]{watson2001}. However, given that these features are reproduced independently in both nights' maps, and given that features with a similar morphology have also been found in AE~Aqr \citep[where there was very dense phase sampling;][]{hill2016}, we interpret these as groups of starspots that have been smeared together in the reconstruction.

\begin{figure}
\includegraphics[width=\columnwidth]{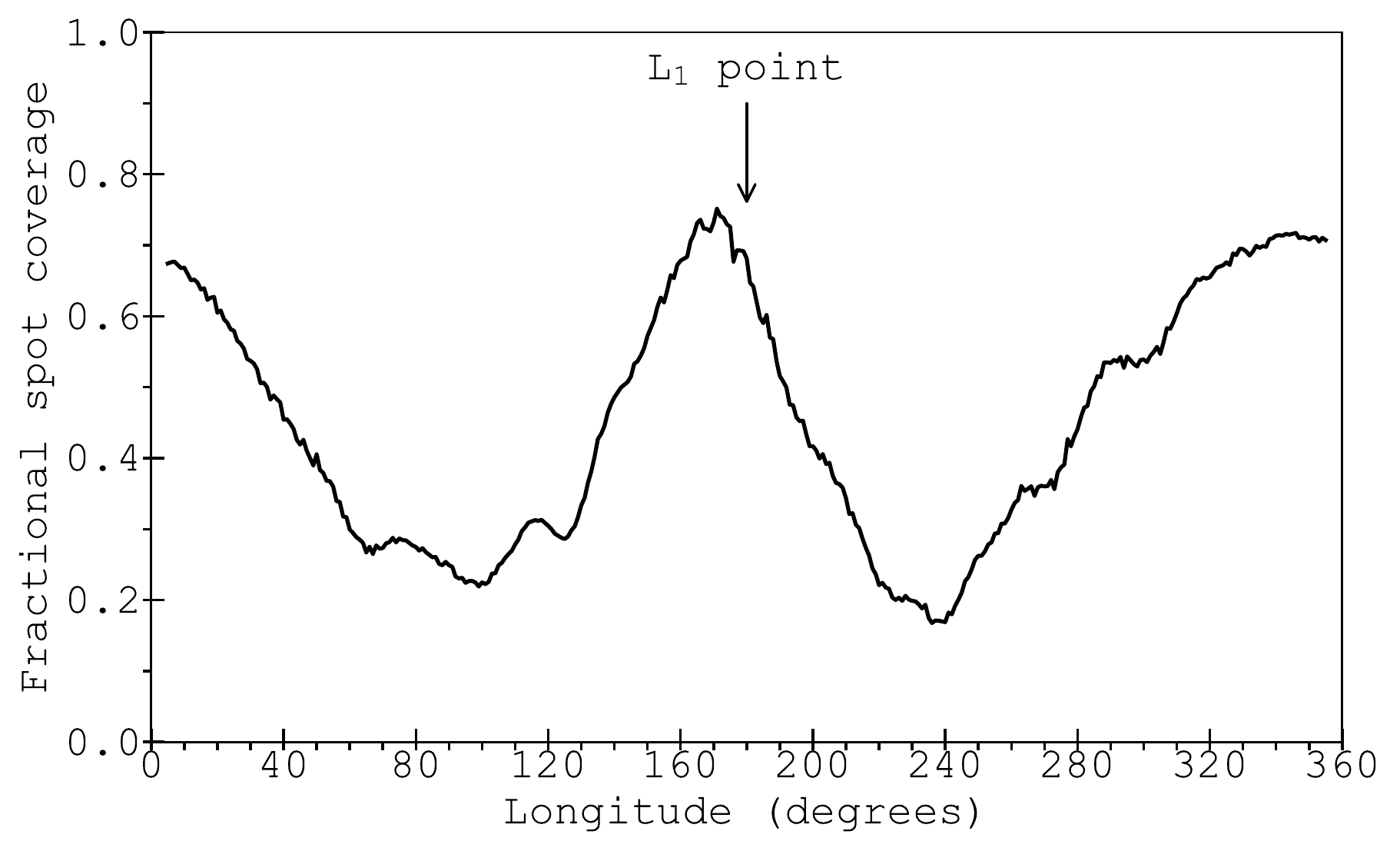}
\caption{Fractional spot coverage as a function of longitude for the map in Figure~\ref{fig:map}, where the first Lagrangian point is at 180\degr.}
\label{fig:long}
\end{figure}

Most prominently on the leading-inner hemisphere of \s, we find a large region affected by irradiation, as is also prominently seen in IP~Peg \citep{watson2003} and RU~Peg \citep{dunford2012}. This brightness asymmetry (stemming from the ionization of atoms, reducing absorption line depth) is clearly seen in Figure~\ref{fig:long}, where the fractional spot coverage peaks around 170\degr (i.e. towards the leading hemisphere). Figure~\ref{fig:long} also shows the increase in spot coverage towards the rear of the star due (predominantly) to the large high-latitude spot.

Following the definition of the immaculate photosphere in \cite{hill2016}, we estimate the total fractional spot coverage for \s to be 39~\pc (for the Northern hemisphere only). Despite systematically underestimating the spot coverage, due to our exclusion of unresolved spots, this value is likely somewhat overestimated, as it additionally includes the effects of irradiation (which appear dark in the maps and are indistinguishable from spots). Indeed, the spot coverage determined here is somewhat larger than the 22~\pc found by \cite{webb2002}; however, it is similar to that of AE~Aqr \citep[28--39~\pc, see][]{hill2016}.

%


\section{Discussion and summary}
\label{sec:discussion}

We have determined the system parameters of \s, and as we take into account the inhomogeneity of the surface brightness on the secondary star, our derived parameters are (in principle) the most robust yet found for this system.

Our derived parameters are given in Table~\ref{tab:systempars}, where we also compare our results to those of other authors. Our systemic velocity of $\gamma = -15.2$~\kms agrees well with the values determined by \citet{hessman1984}, \citet{north2002} and \citet{bitner2007}, and while our optimum value of the radial-velocity semi-amplitude of $\kb = 163.9$~\kms is around 9~\kms higher than that of \citet{hessman1984}, it is in excellent agreement with the values found by \citet{martinezpais1994}, \citet{north2002} and \citet{bitner2007}.
Moreover, our stellar masses of $\ma = 0.94$~\msun and $\mb = 0.59$~\msun, and optimum inclination of $i = 45\degr$ are in excellent agreement with those determined by \citet{giovannelli1983} and \citet{bitner2007}.
Lastly, our optimal mass ratio of $q = 0.628$ is bracketed by the previously measured values of \citet{giovannelli1983}, \citet{hessman1984}, \citet{north2002} and \citet{bitner2007}.


\begin{table*}
\centering
\caption{System parameters of \s, with columns 1--7 listing the authors, the systemic velocity, the inclination, the radial-velocity semi-amplitude of the secondary star, the primary mass, the secondary mass and the mass ratio}
\setlength\tabcolsep{6pt} 
\label{tab:systempars}
\begin{tabular}{lcccccc}
\hline													
Author	&	$\gamma$ (\kms)	&	$i$ (degrees)	&	\kb	&	$M_{1}$ (\msun)	&	$M_{2}$ (\msun)	&	$q = \frac{M_{2}}{M_{1}}$	\\
\hline													
This work	&	$-15.2$	&	45	&	$163.9$	&	0.94	&	0.59	&	0.628	\\
\citet{hessman1984}	&	$-15.1\pm1$	&	-	&	$155\pm2$	&	-	&	-	&	0.595	\\
\citet{north2002}	&	$-$$13.09 \pm 2.88$	&	-	&	$165 \pm 1$	&	-	&	-	&	$0.68 \pm 0.02$	\\
\citet{bitner2007}	&	 $-13.1 \pm 2.9$	&	45--56	&	$162.5\pm1$	&		$0.81 \pm 0.19$	& $0.55 \pm 0.13$	&	$0.685\pm0.015$	\\
\citet{martinezpais1994}	&	-	&	-	&	$162.5 \pm 3$	&	-	&	-	&	-	\\
\citet{giovannelli1983}	&	-	&	$40^{+1}_{-2}$	&	-	&	$0.97^{+0.14}_{-0.05}$	&	$0.56^{+0.08}_{-0.03}$	&	$0.58^{+0.12}_{-0.10}$	\\
\hline											
\end{tabular}
\end{table*}






By constructing Doppler tomograms of the \ha, \hb, \hg and \hd emission lines, we find an asymmetric accretion disc, showing spiral structures, as well as the expected accretion stream, with most of the emission of \ha and \hb coming from the irradiated secondary star.
	
\s is so far the shortest-period dwarf nova to have been mapped with Roche tomography, and so it is of interest to compare the results with those of longer-period CVs such as AE~Aqr \citep{watson2006,hill2014,hill2016} and BV~Cen \citep{watson2007b}. It is difficult to compare the magnetic field strength and topology in these targets without direct measurements of the magnetic field strength (through Zeeman broadening or spectropolarimetric measurements), and so one cannot easily determine how the dynamo, and resulting field strength and topology, are affected by rotation rate or convective zone depth. However, if one makes the reasonable assumption that starspots trace the emergence of magnetic flux tubes, then one can make some general comparisons. The fractional spot coverage of all three of these CVs are fairly similar (around 20--40~per~cent), despite \s rotating 120~\pc faster than BV~Cen, and 50~\pc faster than AE Aqr. Furthermore, all of these stars show large spotted regions at high-latitudes, with many smaller spots distributed across the entire surface. Thus, it appears that the faster rotation of \s does not have a significant impact on the underlying dynamo mechanism, and the resulting magnetic activity. If spot coverage is linked to surface activity, as expected, this result is consistent with the well-established result that stellar activity, which generally increases with decreasing rotation period, saturates for periods less than a day \citep[e.g.][]{pizzolato2003}. It would be desirable to test this further for even shorter-period CVs with M dwarf companions, using larger telescopes and better and faster-readout echelle spectrographs to obtain the same resolution for fainter systems.

It is also interesting that there is clearly some asymmetry in the fractional spot coverage about phase 0.5, with the maximum coverage being shifted about 10$\degr$ towards the leading hemisphere. This effect is consistent with earlier observations of asymmetry in the heating effects of irradiation \citep{smith1995}, which can be explained by Coriolis forces carrying heated material towards the leading hemisphere \citep{martin1995}. Shorter-period systems would be expected to have stronger irradiation, and therefore a larger asymmetry, which strengthens the case for studying the brightness distribution on their rapidly-rotating cool components.

In summary, this study of the well-known dwarf nova \s has revealed some interesting new features, which suggest that the study of even shorter period systems would be valuable.

\section*{Acknowledgements}
Based on observations obtained with the Apache Point Observatory 3.5-m and 1-m telescopes, which are owned and operated by the Astrophysical Research Consortium. LH acknowledges support from NSF grant AST-1009810 and PS acknowledges support from NSF grant AST-1514737. PS is grateful for help from Jon Holtzman (New Mexico State University), who took the 1-m observations, and from Adam Ritchey (University of Washington) for help with using the echelle spectrograph. University of Washington undergraduates Meagan Albright, Caitlin Doughty and Kelsey Braxton took the Manastash Ridge Observatory observations. We thank Christopher Watson for the use of his Roche tomography code, and we thank Tom Marsh for the use of his \textsc{molly} software package, as well as the \textsc{python} version of his Doppler tomography code. This work has made use of the VALD database, operated at Uppsala University, the Institute of Astronomy RAS in Moscow, and the University of Vienna \citep{ryabchikova2015}. 

\bibliographystyle{mnras}
\bibliography{ss_cyg}

\appendix
\label{app:a}

\section{Roche tomograms for each night}
	
\begin{figure*}
\includegraphics[width=0.45\textwidth]{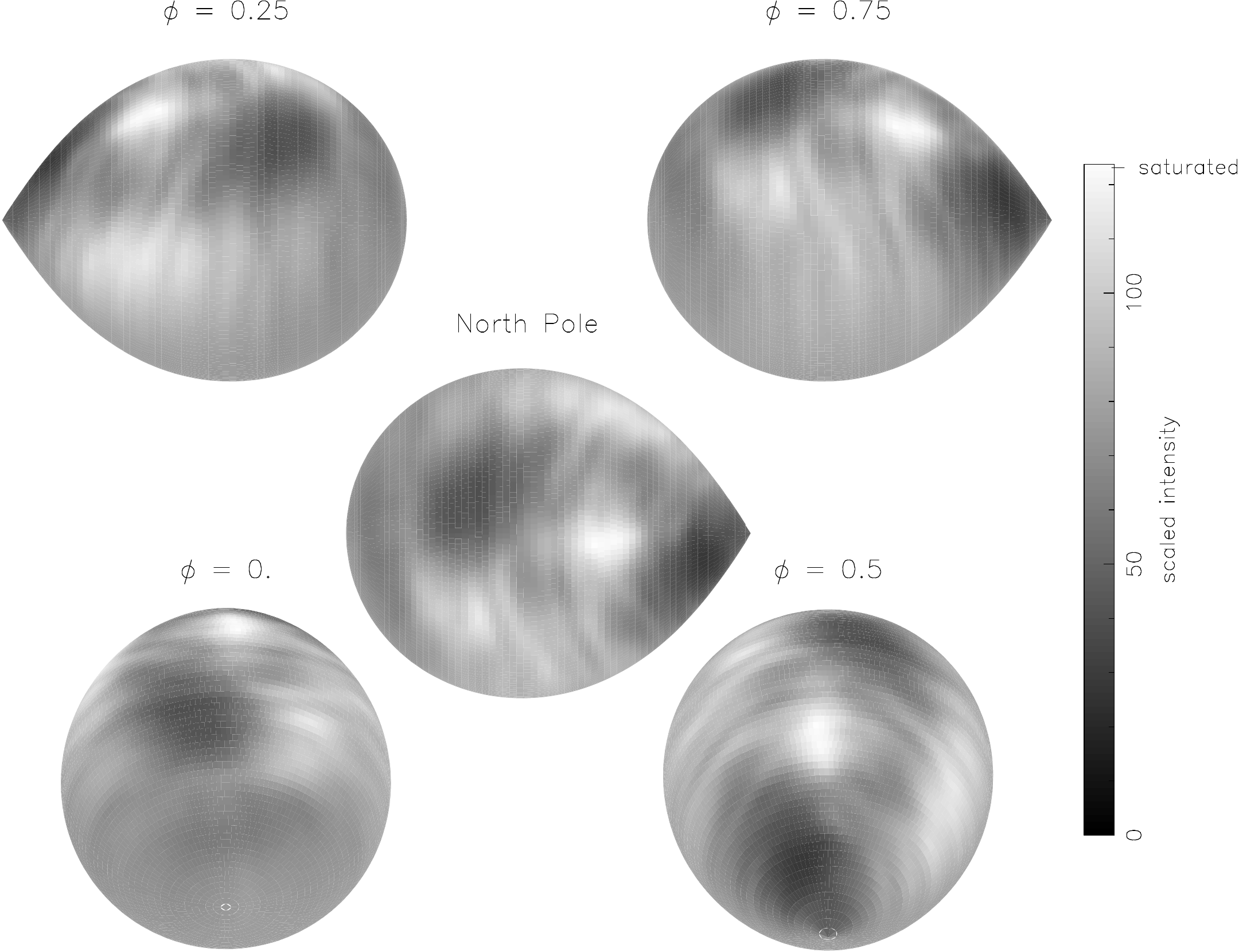}
\includegraphics[width=0.45\textwidth]{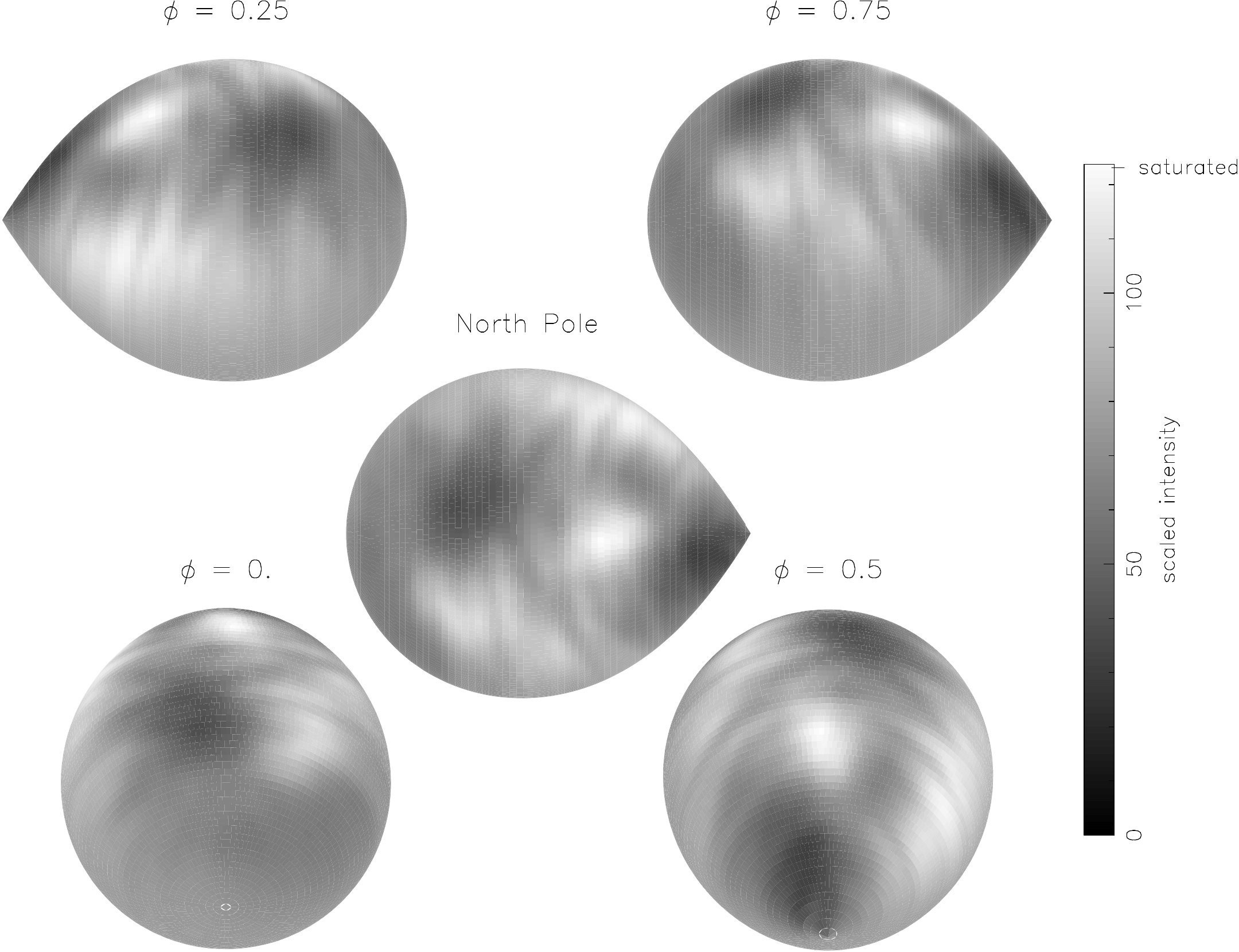}
\caption{Surface maps for the two separate nights. Left panel: night 1; Right panel: night 2. The main features are essentially identical on the two nights.}
\label{fig:maps}
\end{figure*}

In Section~\ref{sec:surfacemaps}, a surface map based on all the data was presented. For completeness, we present here in Figure~\ref{fig:maps} the separate maps for the two nights of observation.

\bsp	
\label{lastpage}
\end{document}